\documentstyle[12pt]{article} 

\begin{document}
\begin{center} 
{\bf {\large {Strange quark matter in cosmic ray flux and exotic events }}} 
\vskip 0.2in 
Shibaji Banerjee$^{\$,\dagger}$, Sanjay K. Ghosh$^{\$,\P,}${\footnote{Present
address: Theory Group, TRIUMF, 4004 Wesbrook Mall, Vancouver, BC V6T 2A3,
Canada}}, Amal
Mazumdar$^{\$,\ddagger,}${\footnote{Permanent address : Fachbereich Physik,
Philipps Universit\"at Marburg, D-35032 Marburg, Germany}}, \\
Sibaji Raha$^{\$,\S,}${\footnote{Corresponding author}} and Debapriyo
Syam$^{\star}$.
\end{center} 
$^{\$}$ Physics Department, Bose Institute, 93/1, A.P.C. Road, Calcutta 
700009, India \\
$^{\star}$ Physics Department, Presidency College, 86/1, College Street, 
Calcutta 700073, India \\ 
$^{\dagger}$ Electronic address : phys@bosemain.boseinst.ernet.in \\
$^{\P}$ Electronic address : sanjay@triumf.ca \\
$^{\ddagger}$ Electronic address : mazumdar@mailer.uni-marburg.de \\
$^{\S}$ Electronic address : sibaji@bosemain.boseinst.ernet.in 
\vskip 0.2in 
\newpage
\begin{abstract}
There have been several reports of exotic nuclear fragments, with highly
unusual charge to mass ratio, in cosmic ray experiments. Although there
exist experimental uncertainties which make them, at best, only candidate
"exotic" events, it is important to understand what they could be, if they
are eventually confirmed. Among other possible explanations, some authors
have interpreted them to be lumps of strange quark matter (strangelets).
A major problem with such an interpretation is that to reach the earth's
surface, they must possess an unusually high penetrability through the
terrestrial atmosphere. We show that a recently proposed mechanism for
the propagation of strangelets through the earth's atmosphere, together
with a proper account of charge capture and ionisation loss, would solve
this problem. We also argue that this could lead to viable strategies for
definitive detection of strange quark matter in cosmic ray flux using a
ground based large area array of passive detectors.
\end{abstract} 
\newpage
There exists a proposal in the literature \cite{witten} that
strange quark matter (SQM), consisting of approximately equal numbers of up,
down and strange quarks, represents the true ground state of Quantum
Chromodynamics (QCD), the underlying theory of strong interaction physics.
The characteristic feature of stable (or metastable) small lumps of SQM
would be very abnormal electric charge ($Z$) to mass ($A$) ratio ( $ Z/A
\ll 1 $). The existence of such objects had also been postulated earlier by
other authors (see, {\it eg} \cite{terazawa}), but the seminal work of Witten
\cite{witten} provided the theoretical basis for the study of SQM 
within the framework of QCD. It has since been argued in the literature that
a definitive confirmation of the existence of stable (or metastable) lumps
of SQM (referred to in the following as strangelets) can shed light on some
of the most intriguing aspects of present day physics and astrophysics, like
the cosmological dark matter problem \cite{alam1}, cosmological QCD phase
transition or abundance of strange stars\cite{madsen1}.
 
There have indeed been several reports of events with $A \sim $ 350 - 500 and 
$Z \sim $ 10 - 20 in cosmic ray experiments \cite{ex1,ex2,ex3}, the so-called
exotic cosmic ray events. Although these observations come from different
groups, the existence of such objects cannot yet be taken as confirmed, due
to various experimental uncertainties like  switch between gondolas,
ambiguities associated with the calibration of Cerenkov counter output,
detector noises, dead time etc, in the different experiments{\footnote{We
thank the referee for emphasizing this point to us.}}. These events, thus,
are, at best, candidate events. Nonetheless, it is an important task to
understand what these objects, if they are eventually confirmed, could be
and several authors have put forward various suggestions as to the nature of
these exotic objects. An essential feature of these objects appears to be
their unusual penetrability through the terrestrial atmosphere, allowing
them to reach mountain altitudes. To account for such stability, although in
the light of a different cosmic ray event called the "Centauro event" found
in the Brazil-Japan collaboration experiment at Mt. Chacaltaya
\cite{centauro}, Bjorken and McLerran \cite{bj} assumed, following
De R\'ujula, Giles and Jaffe \cite{DGJ}, the liberation of a fractionally
charged free massive quark which would absorb a large number of nucleons and
constitute a metastable blob of superdense quark matter. Chin and Kerman
\cite{chin} proposed the existence of metastable multiquark states of large
strangeness within the framework of the MIT bag model \cite{bag}. All such
scenarios can be mostly accommodated within the premise of Witten's
conjecture of SQM as the {\it true} ground state of QCD. However, the
fraction of such heavy objects (A $\sim$ 300 - 500) in the primary cosmic
ray flux may be exceedingly small.
 
It should be mentioned at this juncture that only strangelets with very 
large $A$ were initially thought to be favourable in the context of
stability. Indeed, De R\'ujula and Glashow \cite{DG} considered the
possibility of detecting large lumps of SQM, called "nuclearites", of
$ A   < $ 10$^{15}$ and $Z$ "well beyond any published periodic table".
Recent calculations, however, have shown that also small strangelets with
$A$ = 6, 18, 24, 42, 54, 60, 84, 102 etc are possible stable configurations,
due to an underlying shell-like structure \cite{jaffe,munshi}. The flux of
such objects in cosmic rays could thus be sizable enough to be looked for,
at least in large area detectors, hence ground based experiments. It is thus
imperative to know if these strangelets can traverse the earth's
atmosphere and reach the surface. Moreover, the candidate events ($A \sim$
350 - 500 and $Z \sim$ 10 -20) have vastly different charge-to-mass ratios
than those referred to above. It needs to be investigated whether these are
related in any manner.
 
Recently, a dynamical scenario for the propagation of strangelets through
the earth's atmosphere has been worked out \cite{jpg}, where the
stability of SQM plays a very important role. In particular, the propagation 
of strangelets through the earth's atmosphere has been described by the 
differential equation 
\begin{equation} 
{d{\vec{v}} \over {dt}} = - {\vec{g}} + {\frac{q}{m_s}} ({\vec{v}} \times 
{\vec{B}}) - {\frac{\vec{v}}{m_s}} {{dm_s} \over {dt}} \label{eq:eq1} 
\end{equation} 
where a strangelet of low mass ($A$ = 64 amu) and charge ($Z$ = 2 units) 
enters the upper layer of the atmosphere ($\sim$ 25 km from the sea level, 
above which the density of the atmosphere is negligibly small). The speed of 
the strangelet at that altitude has to be $\geq 0.2c$ ($c$ being the velocity 
of light), for a geomagnetic latitude of $30^o N$, in order to overcome 
the geomagnetic barrier. While the first two terms in equation (\ref{eq:eq1}) 
have obvious significance, the third term accounts for the deceleration of 
the strangelet due to its peculiar interaction with the air molecules; 
strangelets can readily absorb matter and become more strongly bound, unlike 
the normal nuclear fragments which tend to break up \cite{witten}. Using 
straightforward geometrical considerations, it has been shown \cite{jpg} that 
the strangelet grows from $A$ = 64 amu to $A \sim$ 340 amu by the time it 
reaches an altitude of $\sim$ 3.5 km, the altitude of a typical mountain 
peak with adequate accessibility for setting up a large detector array. 
This remarkable possibility makes it imperative to explore the consequences 
of this novel mechanism with greater care, especially taking proper account 
of not only accretion of mass but also that of charge as well as the
dissipation of energy due to ionisation loss.
 
Thus the operative equation (\ref{eq:eq1}) becomes modified to 
\begin{equation} 
{d{\vec{v}} \over {dt}} = - {\vec{g}} + \frac{q}{m_s} ({\vec{v}} \times 
{\vec{B}}) - {\frac{{\vec{v}}}{m_s}} \left( {dm_{sn} \over {dt}} 
+{dm_{sp} \over {dt}}\right) - {\frac{f(v)}{\sqrt{3}{m_s}}} \hat{v}
\label{eq:eq2}
\end{equation} 
where ${\frac{dm_{sn}}{dt}}$ $\left( {\frac{dm_{sp}}{dt}} \right) $ denotes
the accretion to the strangelet due to its interaction with neutrons
(protons) of the air (primarily $N_2$ molecules). It should be noted that
absorption of neutrons would lead only to mass increase while that of
protons would increase both mass and charge of the strangelets.
In equation 2, ${\frac{dm_{sn}}{dt}}$ is related to (${\frac{dm_{sp}}{dt}}$) 
by the following relation, 
\begin{equation} 
\frac{dm_{sp}}{dt}=\frac{\sigma_p}{\sigma_n}{\frac{dm_{sn}}{dt}} 
\equiv f_{pn}\frac{dm_{sn}}{dt} 
\end{equation} 
where $\sigma_n$ and $\sigma_p$ represent the cross sections for the 
absorption of the neutron and the proton, respectively, by the strangelet.
Thus, $f_{pn}$ determines the relative probability for a proton to undergo
the above process vis-a-vis a neutron, and is less than one, on account of
the coulomb barrier present at the surface of the strangelet. The factor
$f(v)$ represents the rate of energy loss due to ionisation of the
surrounding medium by the positively charged strangelet \cite{ion1,ion2}.
The rate of absorption of protons by the strangelet, given by eq.(3),
determines the rate of change in the charge $q$ of the strangelet.
 
The set of equations is solved numerically, using the 4th order Runge-Kutta 
method. The results are shown in Figs.~1 and 2. As can be readily seen from 
Fig.~1, the initial strangelet of $A$ = 64 and $Z$ = 2 evolves into a state
of $A \sim$ 455 and $Z \sim$ 14, very similar to what have been reported in
the literature.
 
Special attention should be paid to Fig.~2. It is found that ionisation loss 
leads to a considerable dissipation of energy and consequent slowdown. It is 
however most interesting to note that at altitudes $\sim $ 3.5 km from the 
sea level, the strangelet has a velocity of the order of 0.0063c, 
corresponding to a total energy of $\sim$ 8.5 MeV. For the present scenario,
this energy (which corresponds to $dE/dx \simeq $ 2.35 MeV/mg/cm$^2$ in the
passive solid state nuclear track detector CR-39), although very small, is
just above the threshold of detection with CR-39, for which
${\left. dE/dx \right|}_{critical}$ turns out to be around 1 MeV/mg/cm$^2$.

The experimental verification of SQM in cosmic ray flux (and the mechanism
of their propagation through the earth's atmosphere) is thus possible with a
suitable ground based detector set up at high altitudes of about 3 to 5 km.
At such altitudes, the predicted energy range of the resulting penetrating
particles with mass $M$ between 300 and 400 and $Z$  between 10 and 15
should lie between 5 to 50 MeV. (This estimate corresponds to an averaging
over all angles of incidence at the top of the atmosphere, taken to be 25
km here, as mentioned above. The estimate of flux would be strongly affected
by the angle of incidence; this is currently under investigation.)
A suitable locality for such observations at an altitude of about 3.5 km
above the sea level has been identified at Sandakphu, in the middle ranges
of the eastern Himalayas, with adequate accessibility and climatic
conditions. Continuous exposure for months or years at a stretch of a
detector assembly with stacks of SSNTDs like CR-39, covering a total area
of about 400 m$^2$, is planned there. (The number of events due to
strangelets may be as few as 5 - 10 per  100 m$^2$ per year, according to
our approximate estimates.) The major considerations in this respect are cost, structural
simplicity, and long time stability of the detection sensitivity against
temperature fluctuations of several tens of Celsius degrees between summer
and winter months and the ruggedness of the passive detectors. Regarding all
these aspects, commercially available CR-39 appears to be the most suitable
choice, which has been shown in NASA SKYLAB experiments \cite{chan,beaujean}
to be capable of detecting heavy ions with energies upto 43 MeV/u. The
signatures produced in such detectors in terms of mass, charge and energy of
detectable strangelets can be evaluated in the expected $dE/dx$ range by
measurements of track dimensions. For this purpose, additional calibration
experiments, exposing CR-39 samples to heavy ions with variable charge
states at almost similar energy ranges, are necessary which can be made at 
several existing heavy ion accelerator facilities. With efficient etching
and automated track measurements, backgrounds of low energy secondary
radiation with lower charge or mass are not expected to pose any serious
problems. Due to specific inherent technical problems like "fading" of
thermoluminescent materials over a long interval of time, they do not seem
to be practical in our experimental conditions. CR-39 has an additional
advantage over the other types of passive semi-conductor detectors using
co-polymers like SR6, CN85 or Lexan; a large amount of characteristic
experimental data are already available for CR-39 in the existing literature.
As alternatives, Mica or Overhead Transparency Foils may also be considered
and calibration experiments using these materials will be conducted at
accelerator facilities to judge their suitability. Other active detectors
and devices do not appear to be suitable for installation at proposed
mountain heights for stand alone operation over long periods and are
therefore not being considered at present.
 
We thus conclude by arguing that detection of strangelets in the cosmic ray 
flux is quite possible, using ground based large area passive detectors at 
mountain altitudes.

{\bf Acknowledgment :} 
The works of SB and SKG were supported in part by the Council of Scientific 
and Industrial Research, Govt. of India, New Delhi. 
\newpage 
\noindent 
{\bf Figure Caption :}\\ 
Figure 1. The variation of mass $m_s$ (in amu) and charge $q$ of the
strangelet with altitude (in km). 
\par 
\noindent 
Figure 2. The variation of energy  $E$ (in MeV) and velocity $\beta$ 
with altitude (in km). The inset shows a zoomed view of the graphs 
for altitude between 5 to 3 km. 
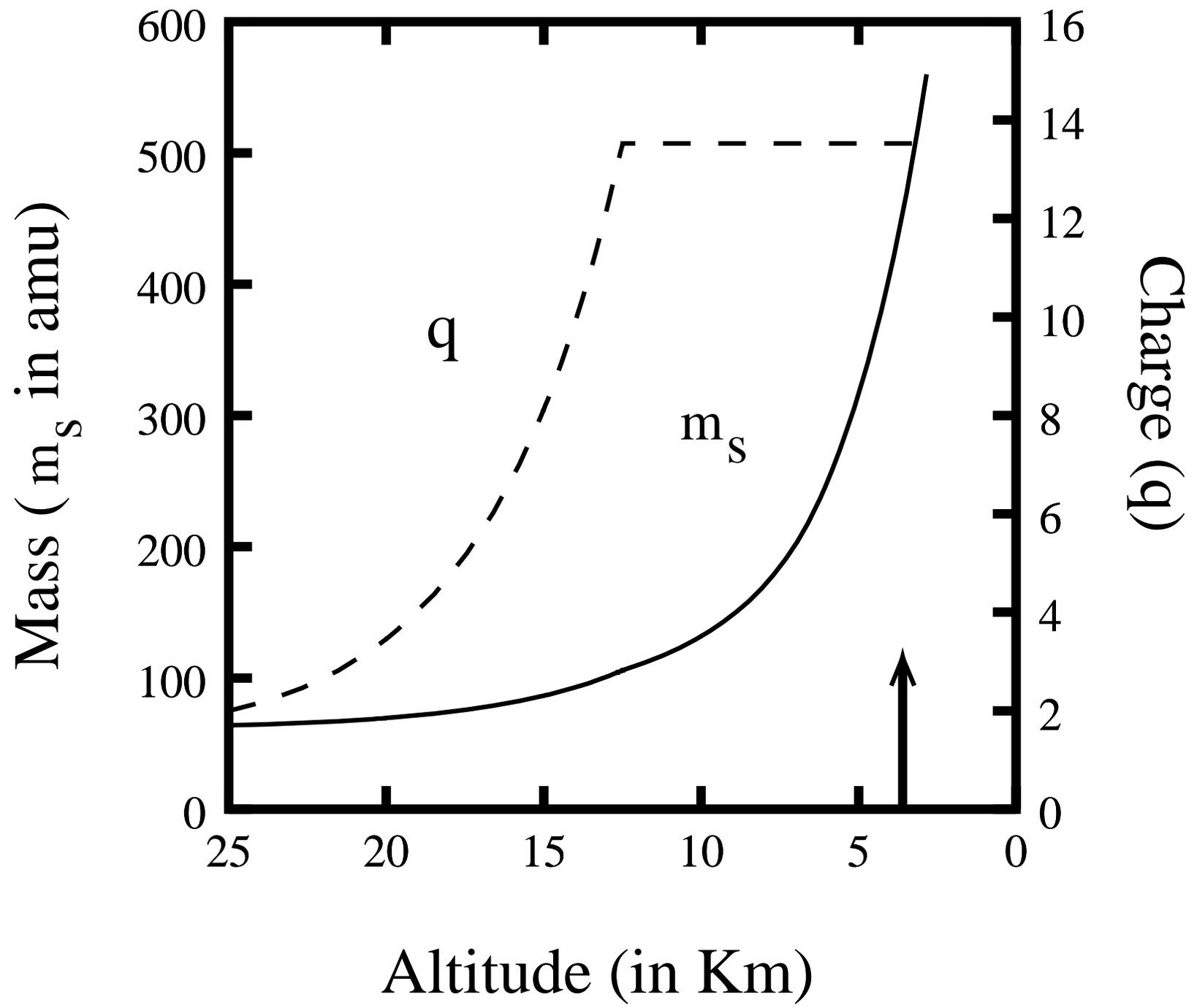
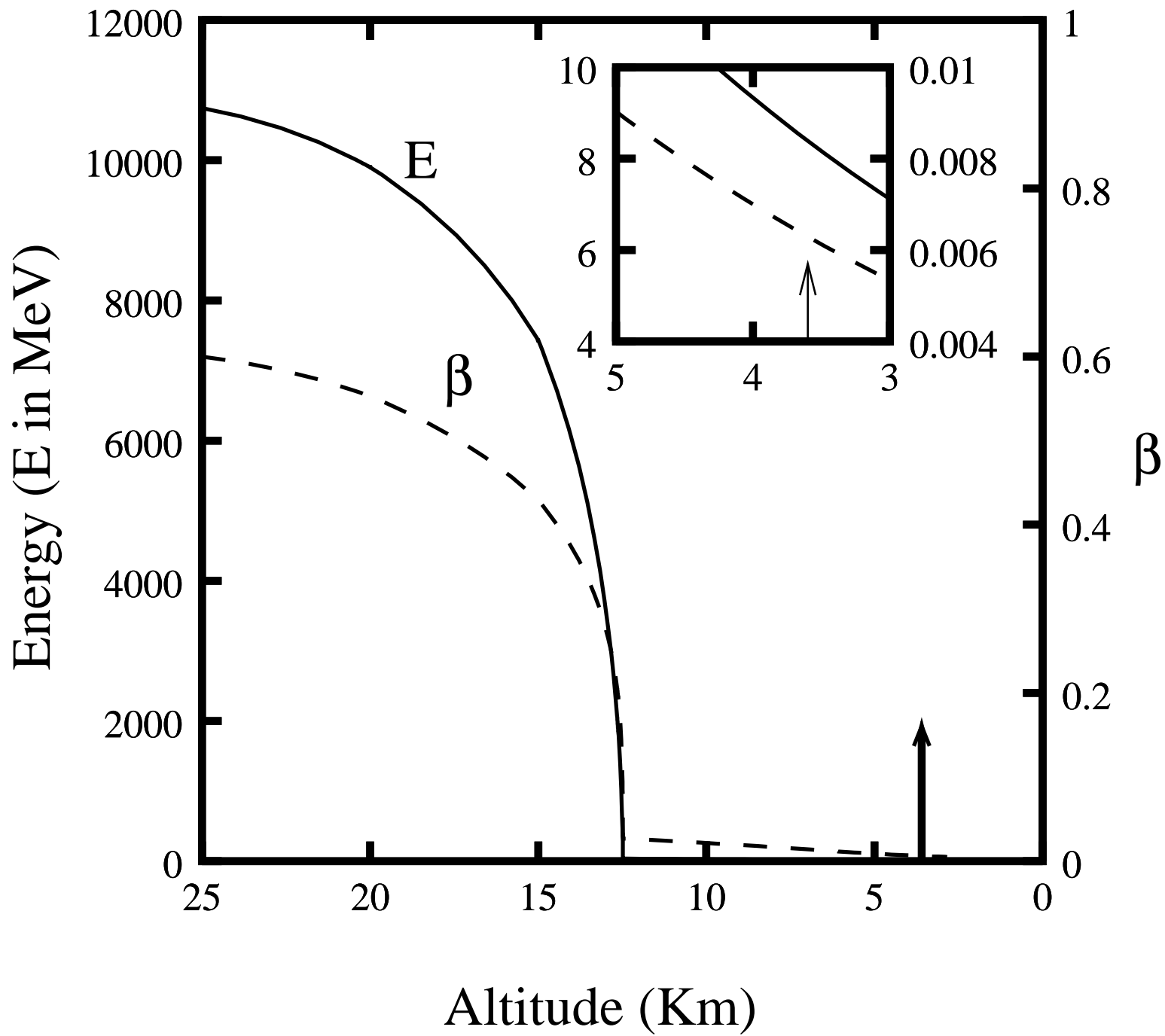
\end{document}